\documentclass[fleqn]{PoS}

\usepackage{newtxtext,newtxmath}
\usepackage{amsmath}

\usepackage{bm}
\usepackage{amssymb}
\usepackage{mathtools}
\usepackage{physics}
\usepackage[mathscr]{euscript}
\usepackage{graphicx}
\usepackage{tabu}
\usepackage{float}
\usepackage{datetime}
\usepackage{multicol}
\usepackage{slashed}
\usepackage{feynmp-auto}
\usepackage[font=footnotesize,labelfont=bf]{caption}
\usepackage{color}
\usepackage{simplewick}
\usepackage{verbatim}
\usepackage{perpage}
\usepackage{setspace}
\usepackage[bottom]{footmisc}

\DeclareMathAlphabet{\mathbbm}{U}{bbm}{m}{n}

\usepackage{subfig}

\newcounter{tmpcount} 
\newcounter{continuemain}
\newcounter{continuesub}

\newcommand {\mathsym}[1]{{}}
\newcommand {\unicode}[1]{{}}

\newcommand{\be}{\begin{equation}}
\newcommand{\ee}{\end{equation}}

\DeclareRobustCommand{\uvec}[1]{{%
  \ifcat\relax\noexpand#1%
    \bm{\hat{#1}}%
  \else
    \ifcsname uvec#1\endcsname
      \csname uvec#1\endcsname
    \else
      \bm{\hat{\mathbf{#1}}}%
     \fi
   \fi
}}

\title{Renormalization of CP-Violating Pure Gauge Operators in Perturbative QCD Using the Gradient Flow}

\ShortTitle{Renormalization with the Gradient Flow}

\author{\speaker{Matthew D. Rizik},\ Andrea Shindler\\
        Facility for Rare Isotope Beams, Physics Department, Michigan State University, East Lansing,
Michigan 48824-4538, USA\\
        E-mail: \email{rizik@nscl.msu.edu}}

\author{Christopher Monahan\\ 
        Institute for Nuclear Theory, University of Washington, Seattle, Washington 98195-1550, USA}

\abstract{We use the Yang-Mills gradient flow to study the mixing of CP-violating pure gauge operators in continuum QCD with special attention to Weinberg's d=6 purely gluonic operator. The gradient flow allows for a relatively clear derivation of the Wilson coefficients of the CP-violating effective Hamiltonian. This calculation is the first step towards a high-energy matching of matrix elements involving the CP-violating operators between the perturbative and lattice regimes.}

\FullConference{The 36th Annual International Symposium on Lattice Field Theory - LATTICE2018\\
		22-28 July, 2018\\
		Michigan State University, East Lansing, Michigan, USA.}

\begin{document}

\section{Introduction} \label{Motivation for the Yang-Mills Gradient Flow}
	The search for a neutron electric dipole moment (nEDM) has become an attractive route for investigating sources of matter-antimatter asymmetry.  The presence of a permanent dipole moment violates both parity (P) and time-reversal (T) symmetry or, equivalently, CP symmetry. The Standard Model of particle physics places a lower bound on the nEDM of $d_n\sim10^{-31}e\cdot\text{cm}$ \cite{Dar:2000tn} from the elecrtoweak sector, while the current supremum is $3.0\times10^{-26}e\cdot\text{cm}$ \cite{Afach:2015sja}, which leaves five orders of magnitude unprobed. To explore this range, the calculation of nucleonic correlators on the lattice may be supplemented with effective sources of CP-violation, higher-dimensional operators coming from effects beyond the Standard Model, an example of which will be briefly addressed in section \ref{The Weinberg Three-Gluon Operator}. However, the mixing of these operators with lower dimensional operators can obscure the determination of the renormalized matrix elements, especially toward the continuum limit. The Yang-Mills gradient flow provides a workaround. It generates a tunable smoothing of the gauge fields by introducing a $(D+1)^{th}$ dimension in which the fields evolve according to the so-called gradient flow equation \cite{Luscher:2010iy}, explained in section \ref{The Yang-Mills Gradient Flow}. This new "flow time" may be infinitesimally varied independently of the lattice spacing, so that the renormalization ($1/\sqrt{8t}$) and regularization ($1/a$) scales disentangle, which should, in principle, conduce to a simpler extraction of the Wilson coefficients for the operator mixing.

\section{Perturbation Theory at Positive Flow-Time} \label{Perturbation Theory at Positive Flow-Time}

	\subsection{Conventions} \label{Conventions}
		We perform all calculations on a CP-even background determined by the standard $SU(N)$-invariant action for interacting fermion fields $\psi(x)$ and gauge fields $A_{\mu}(x)$ in $D$-dimensional Euclidean space:
		\begin{equation}
			S\big[\bar{\psi},\psi,A\big]=\int d^Dx\bigg\{\bar{\psi}(x)\big(\slashed{D}+m_0\big)\psi(x)+\frac{1}{4g_0^2}G_{\mu\nu}^a(x)G_{\mu\nu}^a(x)\bigg\},
			\label{QCDaction}
		\end{equation}
		with bare mass, $m_0$, and coupling, $g_0$. The covariant derivative in the fundamental representation and the curvature are given by
		\begin{equation}
			D_{\mu}=\partial_{\mu}+A_{\mu}^at^a,\hfill G_{\mu\nu}^at^a=G_{\mu\nu}=\big[D_{\mu},D_{\nu}\big],\hfill
			\label{CovDer+Curv}
		\end{equation}
		where the generators, $t^a,\ a=1,\dots,N^2-1$, are chosen such that
		\setcounter{continuemain}{\value{equation}}
		\begin{subequations}
			\begin{equation}
				\big[t^a,t^b\big]=f^{abc}t^c,\hfill \big\{t^a,t^b\big\}=-\frac{1}{N}\delta^{ab}+id^{abc}t^c,\hfill \text{Tr}\big\{t^at^b\big\}=-\frac{1}{2}\delta^{ab}.\hfill
				\label{generators}
			\end{equation}
			\setcounter{continuesub}{\value{equation}}
		\end{subequations}
		Here the quadratic Casimir invariants for the fundamental $(F)$ and adjoint $(A)$ representations are given by
		\setcounter{equation}{\value{continuemain}}
		\begin{subequations}
			\setcounter{equation}{\value{continuesub}}
			\begin{equation}
				C_2(F)=\frac{N^2-1}{2N}\mathbbm{1}_F,\hfill C_2(A)=N\mathbbm{1}_A.\hfill
				\label{Casimirs}
			\end{equation}
			\setcounter{continuesub}{\value{equation}}    
		\end{subequations}
		Lastly, we define the Fourier transform of a function $f$ by
		\begin{equation}
			f(x)=\int_{\mathbb{R}^D}d^Dx\ \tilde{f}(x)e^{ipx}.
		\end{equation}

	\subsection{The Yang-Mills Gradient Flow} \label{The Yang-Mills Gradient Flow}
		The Yang-Mills gradient flow is characterized by the augmentation of a gauge theory with an extra dimension $t$, called flow time (with $[t]=-2$), along which some field(s) evolve according to a modified, nonlinear heat equation. The original theory is imposed as a boundary condition for $t=0$. In the case of $SU(N)$ gauge fields \cite{Luscher:2011bx}, this manifests as
		\begin{equation}
			\partial_tB_{\mu}(x,t)=D_{\nu}G_{\nu\mu}[B]+\alpha_0D_{\mu}[B]\partial_{\nu}B_{\nu},\hfill B_{\mu}(x,0)=A_{\mu}(x).\hfill
			\label{FlowEquation}
		\end{equation}
		The first term on the right-hand side of equation \eqref{FlowEquation} is proportional to the gradient of the Yang-Mills action \cite{ZINNJUSTIN1988297}, while the second is added to avoid perturbative complications, specifically the necessity of damping some UV gauge modes~(cf.~eq.~\eqref{Kernel}). The diffusive nature of equation \eqref{FlowEquation} ensures that the gauge fields are driven toward a minimum of the action, so that they become well-defined functionals of the unflowed field, free of UV singularities.

		The flow equation may be formally solved by splitting equation \eqref{FlowEquation} into a linear and remainder part as \cite{Luscher:2011bx}	
		\setcounter{continuemain}{\value{equation}}
		\begin{subequations}
			\begin{equation}
				\partial_tB_{\mu}(x,t)=\partial_{\nu}\partial_{\nu}B_{\mu}+(\alpha_0-1)\partial_{\mu}\partial_{\nu}B_{\nu}+R_{\mu},
				\label{Linear}
			\end{equation}
			\setcounter{continuesub}{\value{equation}}
		\end{subequations}
		where the nonlinear remainder $R_{\mu}$ is
		\setcounter{tmpcount}{\value{equation}} 
		\setcounter{equation}{\value{continuemain}}
		\begin{subequations}
			\setcounter{equation}{\value{continuesub}}
			\begin{equation}
				R_{\mu}=2\big[B_{\nu},\partial_{\nu}B_{\mu}\big]-\big[B_{\nu},\partial_{\mu}B_{\nu}\big]+(\alpha_0-1)\big[B_{\mu},\partial_{\nu}B_{\nu}\big]+\big[B_{\nu},\big[B_{\nu},B_{\mu}\big]\big],
				\label{Remainder}
			\end{equation}
			\setcounter{continuesub}{\value{equation}}    
		\end{subequations}
		\setcounter{equation}{\value{tmpcount}}
		and employing the heat kernel\footnote{For brevity, we use $\int_p=\int\frac{d^Dp}{(2\pi)^D}$.}
		\begin{equation}
			K_{t}(x)_{\mu\nu}=\int_p\frac{e^{ipx}}{p^2}\Big[\big(\delta_{\mu\nu}p^2-p_{\mu}p_{\nu}\big)e^{-p^2t}+p_{\mu}p_{\nu}e^{-\alpha_0p^2t}\Big],
			\label{Kernel}
		\end{equation}
		so that in momentum space we find an iterative solution in $\tilde{B}_{\mu}$:
		\begin{equation}
			\begin{split}
				\tilde{B}_{\mu}^a(p,t)=\ &\tilde{K}_{t}(p)_{\mu\nu}\tilde{A}_{\nu}^a(p)+\sum_{n=2}^{3}\frac{1}{n!}\int_0^tds\ \tilde{K}_{t-s}(p)_{\mu\nu}\int_{q_1}\cdots\int_{q_n}(2\pi)^D\delta^{(D)}\big(p+q_1+\cdots+q_n\big)\\
				&\times X^{(n,0)}\big(p,q_1,\dots,q_n\big)_{\nu\rho_1\cdots\rho_n}^{ab_1\cdots b_n}\tilde{B}_{\rho_1}^{b_1}(q_1,t)\cdots\tilde{B}_{\rho_n}^{b_n}(q_n,t),
				\label{MomSolution}
			\end{split}
		\end{equation}
		with flow vertices $X^{(2,0)}$ and $X^{(3,0)}$:
		\begin{subequations}
			\begin{equation}
				X^{(2,0)}(p,q,r)_{\mu\nu\rho}^{abc}=if^{abc}\big[(r-q)_{\mu}\delta_{\nu\rho}+2q_{\rho}\delta_{\mu\nu}-2r_{\nu}\delta_{\mu\rho}+(\alpha_0-1)(q_{\nu}\delta_{\mu\rho}-r_{\rho}\delta_{\mu\nu})\big],
			\end{equation}
			\begin{equation}
				\begin{split}
					X^{(3,0)}(p,q,r,s)_{\mu\nu\rho\sigma}^{abcd}=\ &f^{abe}f^{cde}(\delta_{\mu\sigma}\delta_{\nu\rho}-\delta_{\mu\rho}\delta_{\sigma\nu})-f^{ace}f^{bde}(\delta_{\mu\nu}\delta_{\rho\sigma}-\delta_{\mu\sigma}\delta_{\nu\rho}).\\
					&+f^{ade}f^{bce}(\delta_{\mu\rho}\delta_{\sigma\nu}-\delta_{\mu\nu}\delta_{\rho\sigma})
				\end{split}
			\end{equation}
		\end{subequations}

\section{The Weinberg Three-Gluon Operator} \label{The Weinberg Three-Gluon Operator}

	Weinberg has pointed out \cite{PhysRevLett.63.2333} that a dominating contribution to the nEDM should be given by a purely gluonic, dimension-six operator,
	\begin{equation}
		\mathcal{O}_W(x)=\frac{1}{6}id_Wf^{abc}G_{\mu\rho}^a(x)G_{\nu\rho}^b(x)G_{\lambda\sigma}^c(x)\epsilon_{\mu\nu\lambda\sigma},
	\end{equation}
	formed by integrating out Higgs bosons and quarks from a three gluon operator. Since this operator is free from suppression by weak mixing angles and light quark masses, it produces a large contribution to the nEDM.

	\subsection{Operator Basis} \label{Operator Basis}
	In order to renormalize the CP-violating operators, we consider their mixing with lower-dimensional operators in an operator-product expansion. For some gauge-invariant operator at positive flow time, we have the following asymptotic expansion:
	\begin{equation}
		\mathcal{O}^R(x,t)=\sum_ic_i(t)\mathcal{O}_i^R(x,0),
		\label{OPE}
	\end{equation}
	where $\mathcal{O}^R(x,t)$ is a renormalized operator at positive flow time, $\mathcal{O}_i^R(x,0)$ are the renormalized basis operators at zero flow time, and $c_i(t)$ are the corresponding Wilson coefficients.
	This small flow-time expansion (SFTE) may be inserted into any correlators containing the operator and any external probe at nonzero separation, and it affords the particular benefit of explicitly tracking the renormalization scale via $1/8t\sim\mu^2$, since the Wilson coefficients must follow \cite{Luscher:2013vga} 
	\begin{equation}
		c_i(t)\underset{t\rightarrow0}{\propto}t^{\frac{1}{2}\big(d_{\mathcal{O}_i}-d_{\mathcal{O}}\big)}g^{p_i}\Big\{1+\mathcal{O}\big(g^2\big)\Big\},
		\label{WilsonAsymp}
	\end{equation}
	where $d_{\mathcal{O}}$ $\left(d_{\mathcal{O}_i}\right)$ is the scaling dimension of the operator $\mathcal{O}$ $\left(\mathcal{O}_i\right)$, $p_i$ is related to the anomalous dimension of the operator, and $g$ is the coupling given at a scale of $1/\sqrt{8t}$. Fortunately, as shown in ref. \cite{Luscher:2011bx}, the bulk field for pure gauge operators requires no renormalization for $t>0$, so, the left-hand side of \eqref{OPE} is simply the flowed operator, and no renormalization constants need be considered.
		
	In these proceedings, we do not wish to produce an exhaustive basis of operators contributing to the SFTE of the Weinberg operator. We rather focus on the topological charge density,
	\begin{equation}
		q(x,t)=\frac{1}{16\pi^2}\text{Tr}\big\{G_{\mu\nu}(x,t)\tilde{G}_{\mu\nu}(x,t)\big\}.
		\label{TCD}
	\end{equation}
	Since it is the only dimension-four operator contributing at one-loop order, it will provide the dominating contribution for $t\rightarrow0$, as according to equation \eqref{WilsonAsymp}, so that
	\begin{equation}
		\mathcal{O}_W(x,t)\underset{t\rightarrow0}{=}c_q(t)[q(x,0)]_R+c_W(t)[{O}_W(x,0)]_R + \dots\,,
		\label{WeinbergOPE}
	\end{equation}
	where the Wilson coefficient of the topological charge density $c_q(t)$ goes as $1/t$.

	\subsection{Extracting the Wilson Coefficients} \label{Extracting the Wilson Coefficients}
	The topological charge density is most readily probed with a two-gluon external state. Since our asymptotic expression for the Weinberg is constructed near $t=0$, and since the energy scale goes as $1/\sqrt{8t}$, the coupling becomes arbitrarily small, and we may consider our theory at next-to-leading order in perturbation theory.
	We define the following correlation functions
	\begin{subequations}
		\be
			\Gamma^{(2g)}_W(z,t) = \Big\langle A_{\mu}(x)\mathcal{O}_W(z,t)A_{\nu}(y)\Big\rangle,
			\label{eq:gammaW}
		\ee
		\be
			\Gamma^{(2g)}_q(z,t) = \Big\langle A_{\mu}(x)q(z,t)A_{\nu}(y)\Big\rangle.
			\label{eq:gammaq}
		\ee
	\end{subequations}
	Inserting the SFTE of the Weinberg operator~\eqref{WeinbergOPE} in the correlation function~\eqref{eq:gammaW} and expanding
	in the coupling constant, we obtain
	\be
       	 	\begin{split}
			\Gamma^{(2g)}_W(z,t)^{(0)} + g_0^2\Gamma^{(2g)}_W(z,t)^{(1)}=&\ \left(c_q^{(0)}(t) +g_0^2 c_q^{(1)}(t)\right)\left( \Gamma^{(2g)}_q(z,0)^{(0)}+g_0^2 \Gamma^{(2g)}_q(z,0)^{(1)} \right)\\
			&\ +\left(c_W^{(0)}(t) +g_0^2 c_W^{(1)}(t)\right)\left( \Gamma^{(2g)}_W(z,0)^{(0)} + g_0^2 \Gamma^{(2g)}_q(z,0)^{(1)} \right) + \cdots\,.
		\end{split}
		\label{eq:NLOWeinbergOPE}
	\ee
	The tree-level correlator $\Gamma^{(2g)}_W(z,t)^{(0)}$ vanishes for every $t$, implying that the leading mixing coefficient is of order $g_0^2$, so
	$c_q^{(0)}(t)=0$. At the next order in the coupling, again equating the same order in both sides of eq.~\eqref{eq:NLOWeinbergOPE}, we find
	\be
		\Gamma^{(2g)}_W(z,t)^{(1)} = c_q^{(1)}(t)\Gamma^{(2g)}_q(z,0)^{(0)} + c_W^{(0)}(t)\Gamma^{(2g)}_q(z,0)^{(1)}\,.
	\ee
	The tree-level value $c_W^{(0)}$ can be set to one after inspecting the tree level of a similar expansion with three external gluons.
	Thus we arrive at the defining equation for the Wilson coefficient:
	\be
		c_{q}^{(1)}(t) = \frac{1}{\Gamma^{(2g)}_q(z,0)^{(0)}}\left[\Gamma^{(2g)}_W(z,t)^{(1)} - \Gamma^{(2g)}_q(z,0)^{(1)}\right]\,.
		\label{powerdivergent}
	\ee
        Once the power-divergent mixing of the Weinberg operator with the topological charge density is calculated, the one-loop Wilson coefficients of other operators may be extracted.
	\begin{figure*}[hbpt]
		\centering
		\subfloat[]{\includegraphics[width=.3\textwidth]{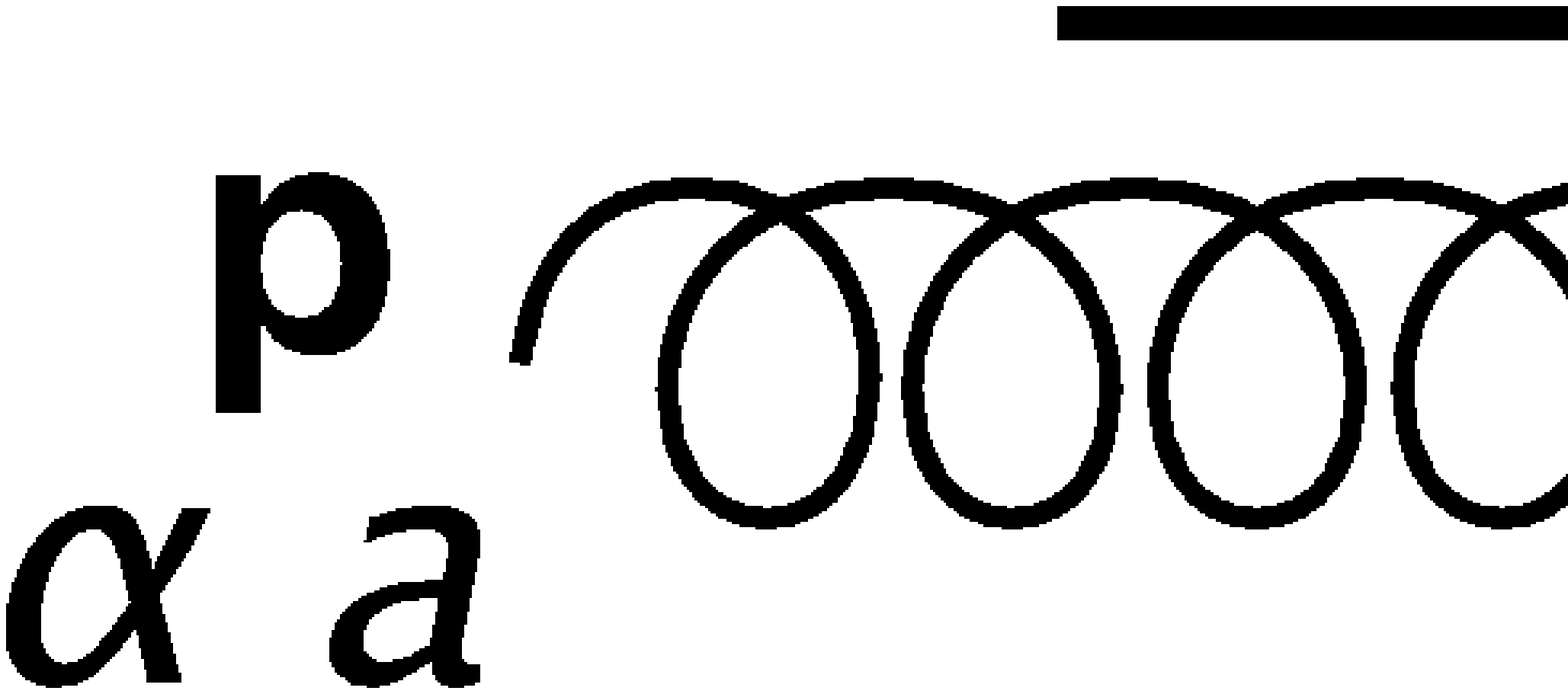}\label{3gv-unflowed}}\,
		\subfloat[]{\includegraphics[width=.3\textwidth]{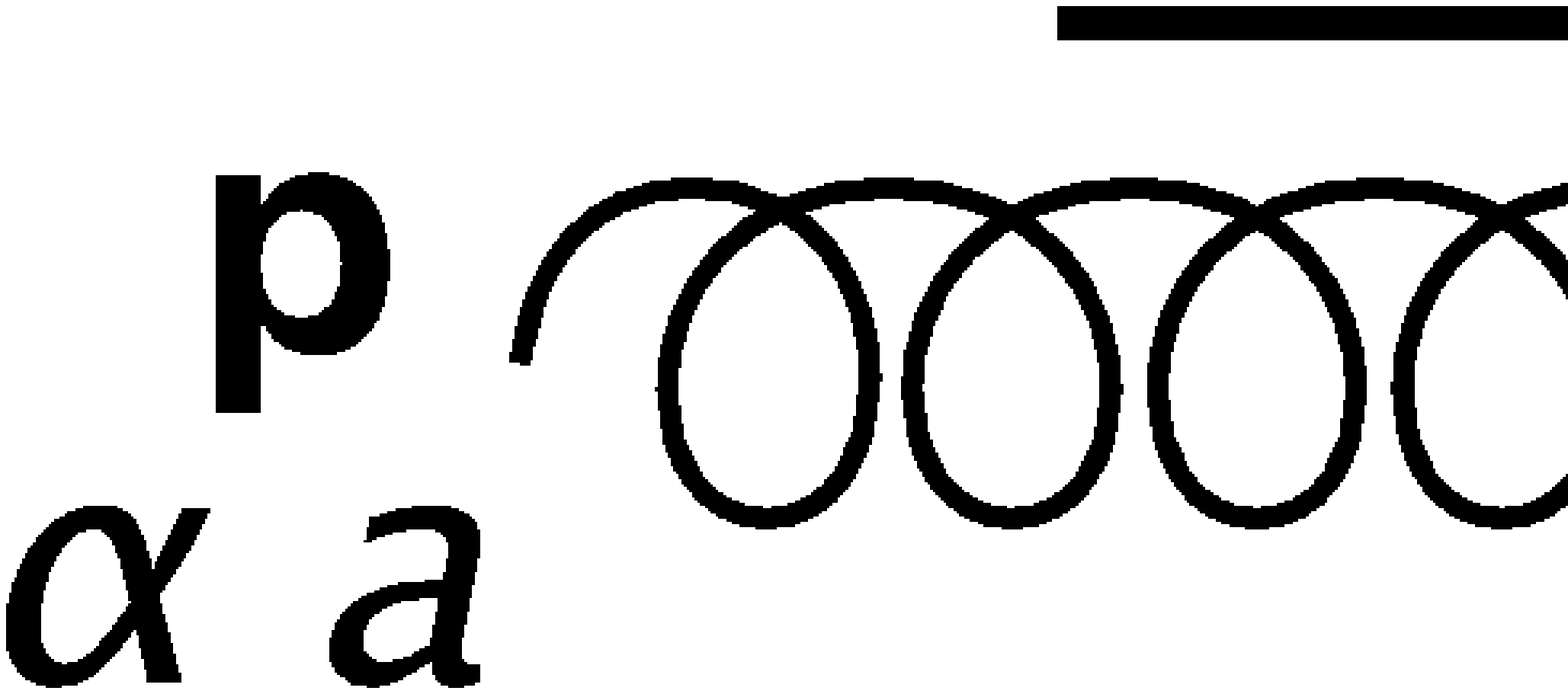}\label{3gv-flowed}}\,
		\subfloat[]{\includegraphics[width=.3\textwidth]{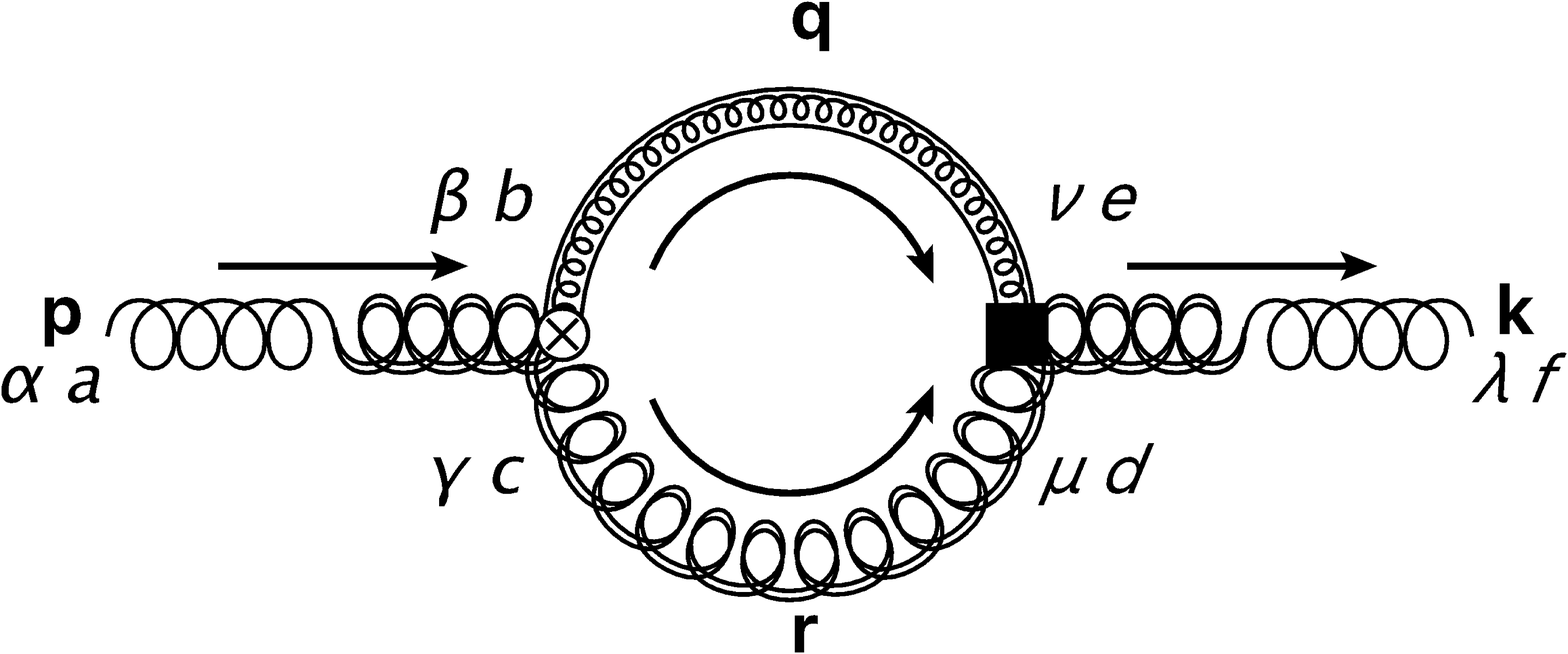}\label{3gv-kernel}}\\
		\subfloat[]{\includegraphics[width=.3\textwidth]{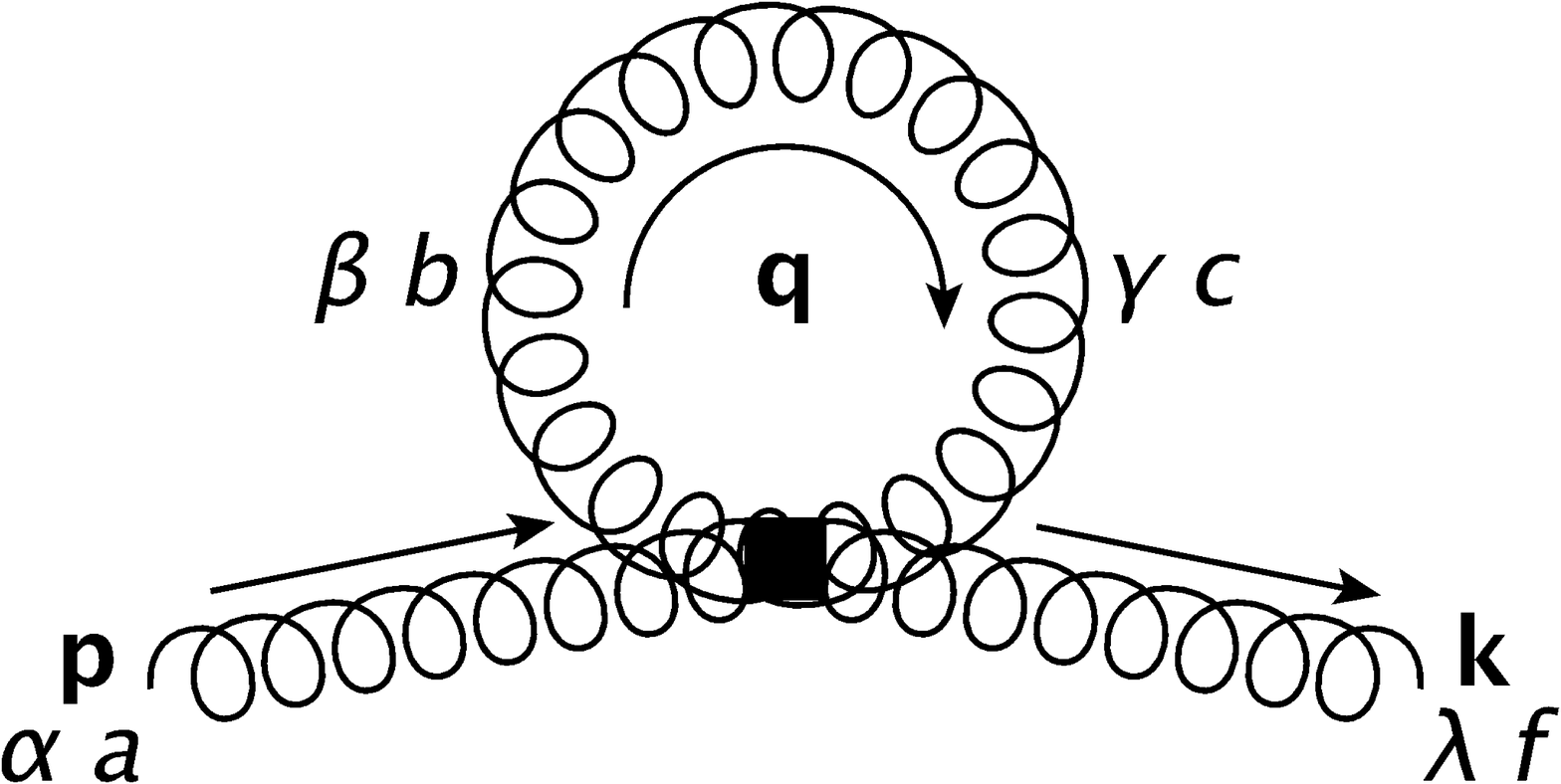}\label{4gv-unflowed}}\,
		\subfloat[]{\includegraphics[width=.3\textwidth]{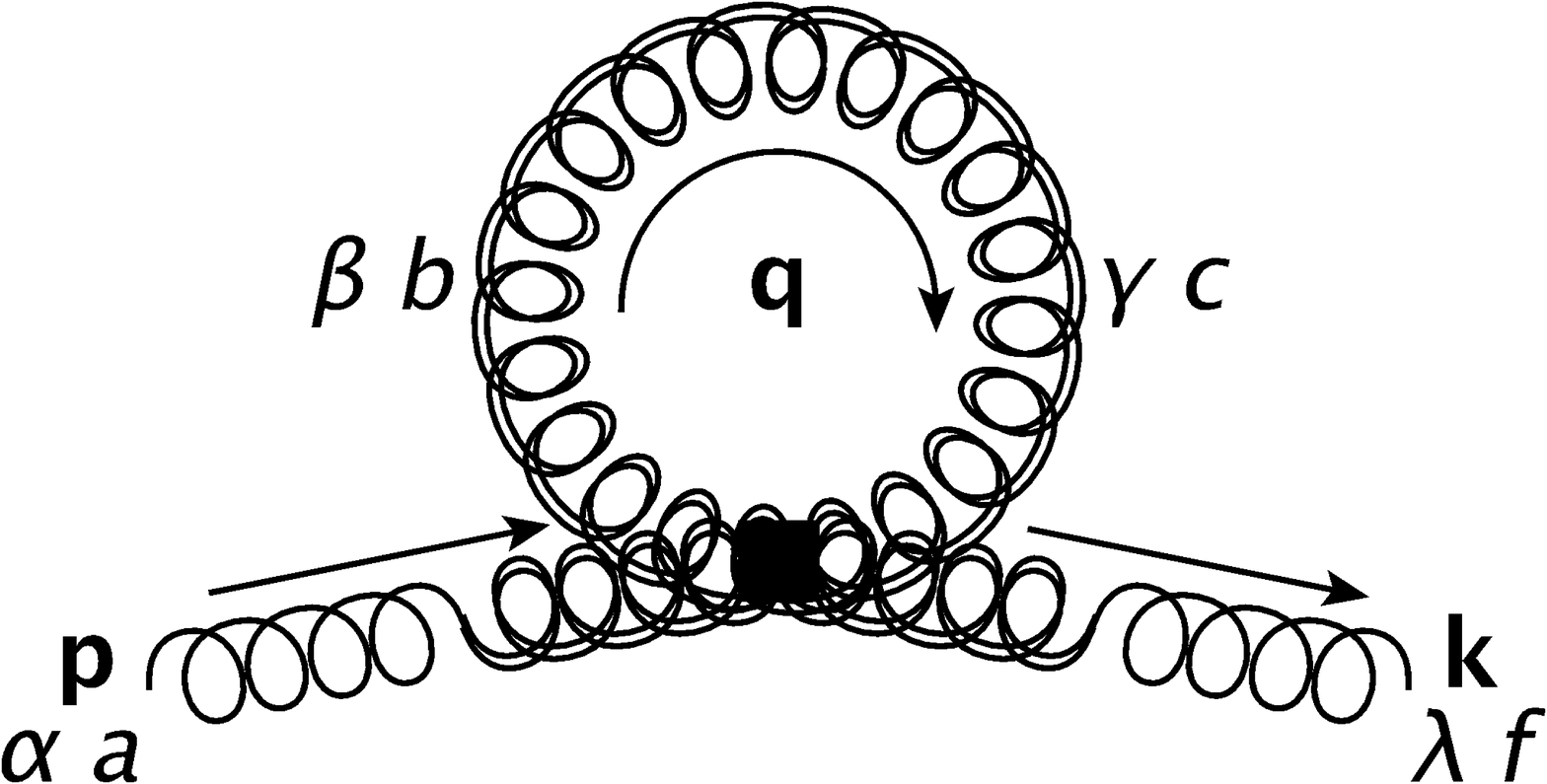}\label{4gv-flowed}}
		\caption{Contributions to the one-loop Wilson coefficient of the topological charge density. The Weinberg operator is represented by a black square, flowed gluons are represented by double curly lines, kernels for any particle species are represented by their standard propagator sandwiched between two solid lines (the kernel transports the particle data without propagation), and flow vertices $X^{(n,0)}$ are represented by crossed open circles. All other notations are standard.}
		\label{}
	\end{figure*}
	    
	The first correlator in the numerator of equation \eqref{powerdivergent} admits three (at least nontrivially) non-vanishing graphs: a simple self-energy diagram (Fig. \ref{3gv-flowed}), a tadpole (Fig. \ref{4gv-flowed}), and a "kernel" diagram (Fig. \ref{3gv-kernel}) resulting from the first iteration of equation \eqref{MomSolution}. In dimensional regularizaton, the general solution of these integrals is unknown for arbitrary external states, since the flowed propagator gives rise to terms without spherical symmetry, which often prohibits the standard analytic continuation over the spherical part of the integral. In practice, however, we need only consider terms which are at most linear in the external momentum $p_{\mu}$, so we may perform a Taylor series expansion about $p_{\mu}$, simply ignoring higher-order terms which do not mix. For example, in diagram \ref{3gv-flowed} we encounter integrals of the form
		\begin{equation}
			I_n^D(p,t)=\int_q\prod_{i=1}^nq_{\mu_i}\frac{e^{-q^2t-(p-q)^2t}}{q^2(p-q)^2}=\int_q\prod_{i=1}^nq_{\mu_i}\frac{e^{-2q^2t}}{(q^2)^2}\bigg[1+2(1+q^2t)\frac{q_{\mu}p_{\mu}}{q^2}+\mathcal{O}\big(p^2\big)\bigg].
			\label{FlowIntegral}
		\end{equation}
		Note that the particular integral on the left above can in principle be solved completely analytically \cite{MonahanRizikShindler}, but this is unnecessary for the mixing calculation. The second term in brackets is odd in the integration momentum, so it generates additional contributions for odd $n$ which would otherwise vanish under integration in standard QCD. This procedure formally requires an infrared regulator $\epsilon<0$, such that $D=4-2\epsilon$, since the infrared regulator $p$ has been brought out of the denominator, though all dependence on $\epsilon$ vanishes in this case.  In this scheme, the diagrams evaluate to
		\begin{subequations}
			\begin{equation}
				\big(\Gamma_{\ref{3gv-flowed}}\big)_{\alpha\lambda}^{af}(p,k,t)=-6\frac{g_0^2}{(4\pi)^2}d_WC_2(A)\tilde{q}^{(0)}(p,k)\frac{1}{t}\Big\{1+\mathcal{O}\big(p^2,t\big)\Big\},
				\label{3flowed}
			\end{equation}
			\begin{equation}
				\big(\Gamma_{\ref{4gv-flowed}}\big)_{\alpha\lambda}^{af}(p,k,t)=0,
				\label{4flowed}
			\end{equation}
			\begin{equation}
				\big(\Gamma_{\ref{3gv-kernel}}\big)_{\alpha\lambda}^{af}(p,k,t)=\frac{g_0^2}{(4\pi)^2}d_WC_2(A)\tilde{q}^{(0)}(p,k)\frac{1}{t}\Big\{1+\mathcal{O}\big(p^2,t\big)\Big\}.
				\label{3kernel}
			\end{equation}
			\setcounter{continuesub}{\value{equation}}
		\end{subequations}
		The second term in equation \eqref{powerdivergent} must be treated similarly. The two contributing diagrams are givens in figures \ref{3gv-unflowed} and \ref{4gv-unflowed}. They are given through standard techniques by
		\begin{subequations}
			\begin{equation}
				\begin{split}
					\big(\Gamma_{\ref{3gv-unflowed}}\big)_{\alpha\lambda}^{af}(p,k,0)=\ &2\frac{g_0^2}{(4\pi)^2}d_WC_2(A)\tilde{q}^{(0)}(p,k)\cdot p^2\bigg\{\frac{1}{\epsilon}-\text{log}\bigg(\frac{\gamma'_Ep^2}{4\pi\mu^2}\bigg)+\frac{8}{3}\bigg\}\\
					\rightarrow\ &\tilde{q}^{(0)}(p,k)\Big\{0+\mathcal{O}\big(p^2,t\big)\Big\},
				\end{split}
				\label{3unflowed}
			\end{equation}
			\begin{equation}
				\big(\Gamma_{\ref{4gv-unflowed}}\big)_{\alpha\lambda}^{af}(p,k,0)=0.
				\label{4unflowed}
			\end{equation}
			\setcounter{continuesub}{\value{equation}}
		\end{subequations}
		Subsequently, we find an expression for the Wilson coefficient of the topological charge density:
		\begin{equation}
			c_{q}^{(1)}(t)=-5\frac{g_0^2}{(4\pi)^2}d_WC_2(A)\frac{1}{t}\underset{N\rightarrow3}{\sim}-5d_Wt\langle E(t)\rangle,
			\label{4flowed}
		\end{equation}
		which confirms the dependence of $c_{q}^{(1)}(t)$ on flow time as expected from equation \eqref{WilsonAsymp}. The last step has used L\"uscher's expression for the energy at positive flow time \cite{Luscher:2010iy}. This form is particularly useful for perturbative matching to lattice data.
	
		The remaining coefficients may be calculated in a similar manner, by choosing other probes.  Since the expanded operator is intrinsically renormalized at positive flow time, the renormalization group evolution of the SFTE is encoded entirely in the Wilson coefficients. When they are all calculated, all but the self-mixing term may be subtracted off, so that the remaining term is proportional to the renormalized operator at zero flow time by construction.  In the case of the Weinberg operator, the constant of proportionality is the anomalous dimension at positive flow time, which may be extracted by studying three-gluon external states.
		
		This procedure is general, so it may be applied to any pure gauge operator. Since the flow does not renormalize fermions, however, there is additional scale dependence of an expanded fermionic operator, and the renormalization is not so straightforward.

\end{document}